\documentclass[a4paper,11pt]{article}
\usepackage{pos}

\title{$\hbar$ -perturbative solutions of quantum Snyder and Yang models with parameters describing
spontaneous symmetry breaking}

\author*[a]{Jerzy Lukierski}
\author[b]{Anna Pacho\l}

\affiliation[a]{Institute of Theoretical Physics, Wroc\l aw University,\\ pl. Maxa Borna 9, 50-205 Wroc\l aw, Poland}

\affiliation[b]{Department of Microsystems, University of South-Eastern Norway,\\ Campus Vestfold, Raveien 215, 3184 Borre, Norway}

\emailAdd{jerzy.lukierski@uwr.edu.pl}
\emailAdd{anna.pachol@usn.no}

\abstract{We introduce the perturbative $\hbar $-power series ($\hbar $ - Planck
constant) providing the algebraic solutions of $D=4$ quantum   Snyder and Yang  models which describe relativistic quantum space-times and Lorentz-covariant quantum phase spaces. We argue that if in these series the zero order ($\hbar $-independent) terms are non-vanishing they describe the spontaneous symmetry breaking (SSB) parameters
of Lie-algebraic symmetries which characterize the considered models ($D=4$ dS
symmetry in Snyder and $D=5$ dS symmetry in Yang cases). The consecutive
terms in $\hbar $-power series can be calculated explicitly if we supplement
the SSB order parameters (Nambu-Goldstone or NG modes) by dual set of commutative
momenta, which together define the canonical tensorial Heisenberg
algebra.}

\FullConference{%
  Corfu Summer Institute 2022 "School and Workshops on Elementary Particle Physics and Gravity"\\
  28 August - 1 October, 2022\\
  Corfu, Greece
}

\begin{document}
\maketitle

\section{Introduction}
It is well known (see e.g. \cite{1,2,3,4,5}) that in quantum theories one can
consider two ways of symmetry breaking. The first, explicit symmetry
breaking, leads to modified symmetry properties of algebraic structure in
considered quantum models (e.g. of action integrals, quantum equations of motion
etc.). The second way, describing the case of spontaneous symmetry breaking (SSB), leads to modified symmetries of the solutions and quantum states obtained as the particular SSB realizations of basic algebraic structure. The SSB effects have been considered in Quantum Mechanics (QM) and Quantum
Field Theory (QFT) models (see e.g. \cite{6Nambu,7Goldstone,8Coleman,9}), in particular in the Standard Model (SM), which describes the theory of elementary particles
by the tools of QFT (see e.g. \cite{9a,9b}). It appears also that in SM the suitable SSB of local gauge symmetries generates, by means
of Higgs mechanism \cite{10,11}, the mass parameters which are necessary for the physical applications of SM. 

In this paper we consider, in the presence of Quantum Gravity (QG), the effects of the SSB of quantum space-time symmetries, and for such a purpose we use the
formalism of noncommutative (NC) geometry, used already since a long time for the
description of passage from classical to quantum gravity. It should be
observed that using the tools of NC geometry (see e.g. \cite{12,13,14,15}), there have been obtained various models (see e.g. \cite{16,17,18,19,20})
describing $D=4$ quantum space-times, quantum deformed phase spaces and quantum
symmetry groups. The NC models which preserved firstly the $D=4$ relativistic
covariance were introduced as early as in 1947 by Snyder and Yang. These models were considered in numerous papers (see e.g. \cite{22,23,24,25,26,21a,21b}), where however the appearance of  SSB effects have not been yet studied. It
should be recalled, however, that in \cite{26,27,28,29} the Snyder type models were
solved perturbatively as embedded in the canonical vectorial and tensorial
Heisenberg algebras \cite{30,31}.
In this paper, we will show that the introduction  of explicit $\hbar$-dependence and the use of perturbation
theory described by $\hbar-$power series permits to provide the SSB
interpretation of the obtained results \footnote{See also \cite{33} where such interpretation of SSB in Snyder and
Yang models has been firstly presented; for the discussion of $\hbar$-power expansions in quantum theories, see e.g. \cite{34}.}.
\section{Quantum $D=4$ Snyder algebra, spontaneous symmetry breaking and degenerate vacua}
Firstly, we outline briefly the algebraic $\hbar$-dependent description of the Snyder model.
The classical relativistic Minkowski space-time coordinates $%
x_{\mu }\in\mathbb{M}^{3,1}$ can be introduced as irreducible vectorial
representation of the $D=4$ Lorentz algebra:
\begin{equation}\label{MM}
[M_{\mu \nu },M_{\rho \tau }]=i(\eta _{\mu \rho }M_{\nu \tau }-\eta _{\mu
\tau }M_{\nu \rho }+\eta _{\nu \tau }M_{\mu \rho }-\eta _{\nu \rho }M_{\mu
\tau }).
\end{equation}
The relativistic Lorentz covariance of $\mathbb{M}^{3,1}$ can be expressed by the
relation:
\begin{equation}
\lbrack M_{\mu \nu },x_{\rho }]=i(\eta _{\mu \rho }x_{\nu }-\eta _{\nu \rho
}x_{\mu }).  \label{Mx}
\end{equation}
The covariance of fourmomenta $p_{\mu }\in \mathbb{P}^{3,1}$ is defined in an
analogous way:
\begin{equation}\label{MP}
\lbrack M_{\mu \nu },p_{\rho }]=i(\eta _{\mu \rho }p_{\nu }-\eta _{\nu \rho
}p_{\mu }).
\end{equation}
The quantum relativistic space-time coordinates $\hat{x}_{\mu }\in \hat{\mathbb{X}}^{3,1}$ and quantum fourmomenta  $\hat{p}_{\mu }\in \hat{\mathbb{P}}^{3,1}$ will be
specified if we supplement the algebraic noncommutativity relations (\ref{MM}-\ref{MP}) by the
set of explicit commutators: $\left[ \hat{x}_{\mu },\hat{x}_{\nu }\right] ,\left[ 
\hat{p}_{\mu },\hat{p}_{\nu }\right] $ and $\left[ \hat{x}_{\mu },\hat{p}%
_{\nu }\right] $.
For further study, we should pass from the classical to quantum Lorentz
algebra by the following rescaling of the generators: 
\begin{eqnarray}\label{Mhat}
\hat{M}_{\mu \nu }=\hbar M_{\mu \nu }
\end{eqnarray}
which leads to the following $\hbar$-dependent Lie-algebraic relations:
\begin{equation}\label{MhatMhat}
\lbrack \hat{M}_{\mu \nu },\hat{M}_{\rho \tau }]=i\hbar (\eta _{\mu \rho }%
\hat{M}_{\nu \tau }-\eta _{\mu \tau }\hat{M}_{\nu \rho }+\eta _{\nu \tau }%
\hat{M}_{\mu \rho }-\eta _{\nu \rho }\hat{M}_{\mu \tau }).
\end{equation}
By using (\ref{Mhat}) we obtain that the
covariance relations (\ref{Mx}, \ref{MP}) take the form:
\begin{equation}\label{Mhatx}
\lbrack \hat{M}_{\mu \nu },\hat{x}_{\rho }]=i\hbar (\eta _{\mu \rho }\hat{x}%
_{\nu }-\eta _{\nu \rho }\hat{x}_{\mu }),
\end{equation}
\begin{equation}\label{Mhatp}
\lbrack \hat{M}_{\mu \nu },\hat{p}_{\rho }]=i\hbar (\eta _{\mu \rho }\hat{p}%
_{\nu }-\eta _{\nu \rho }\hat{p}_{\mu }).
\end{equation}
We specify the Snyder quantum space-time by introducing Snyder NC coordinates $\hat{x}_{\mu }$ satisfying the basic relation 
\begin{equation}\label{snyder_l}
\lbrack \hat{x}_{\mu },\hat{x}_{\nu }]=i{l^{2}}M_{\mu \nu }=i{\frac{l^{2}}{\hbar }}\hat{M}_{\mu \nu },
\end{equation}%
where $l$ is an elementary length. If we assume that $c=1$, by using the Compton wavelength formula (see e.g. \cite{34}), one can replace $%
l$ by the inverse of mass $M$, multiplied by $\hbar $, i.e.
\begin{equation}
l=\frac{\hbar }{Mc}\longrightarrow l=\frac{\hbar }{M}
\end{equation}
In this way we obtain:
\begin{equation}\label{snyder_m}
\lbrack \hat{x}_{\mu },\hat{x}_{\nu }]=i{\frac{\hbar }{M^{2}}}\hat{M}_{\mu \nu }.
\end{equation}

Snyder quantum space-time coordinates, with the Lorentz covariance relation (\ref{Mhatx}) and their NC structure provided by (\ref{snyder_m}), can be described by the curved NC translations spanning the $D=4$ dS coset $o(4,1)/o(3,1)$ as
\begin{equation}\label{M4}
\hat{M}_{4 \mu}=M \hat{x}_\mu,\qquad (\mu=0,1,2,3).
\end{equation}
It follows, from relations (\ref{MhatMhat}), (\ref{Mhatx}), (\ref{M4}) that the generators $\hat{M}_{AB}$ $(A,B=0,1,2,3,4)$ providing the quantum $D=4$ dS algebra 
\begin{equation}
\lbrack \hat{M}_{AB},\hat{M}_{CD}]=i\hbar (\eta _{AC}\hat{M}_{BD}-\eta _{AD}%
\hat{M}_{BC}+\eta _{BD}\hat{M}_{AC}-\eta _{BC}\hat{M}_{AD}).
\label{dSMM}
\end{equation}
describe the quantum $D=4$ Snyder model. 

We recall that the difference between the symmetries of equations of motion and the restricted symmetries of explicit solutions in the model lead to the appearance of SSB. 
Namely, the SSB arises in the model when the symmetries describing covariance of the algebraic structure are reduced due to the presence of numerical order parameters in explicit solutions, as well as in background fields (see e.g. \cite{33a}).
In our case the algebra (\ref{dSMM}) is $D=4$ Lorentz and $D=4$ dS covariant, but the realizations of generators
$\hat{M}_{AB}$ may contain parts which violate both the Lorentz and dS covariance. In the general case, one can decompose $\hat{M}_{AB}$, into their classical and quantum parts, as follows:
\begin{equation}\label{Mcl-q}
\hat{M}_{AB}=M^{cl}_{AB}+\hat{M}_{AB}^{q}=M^{cl}_{AB}+\hbar{M}_{AB}^{q}
\end{equation}
where the classical Abelian part is given by the $5\times 5$ antisymmetric matrices
\begin{equation}\label{Mcl}
M^{cl}_{AB}=\lim_{\hbar\to 0}\hat{M}_{AB}=x_{AB}=-x_{BA}.
\end{equation}
In such a way, we introduce the NG degrees of freedom and the spontaneous symmetry breaking of the quantum $D=4$ dS algebra (\ref{dSMM})\footnote{The non-vanishing classical part $M_{AB}^{cl}$ may define as well the classical background parameters which lead to spontaneously broken symmetries.}.
We see that, if we describe $\hat{M}_{AB}$ in (\ref{Mcl-q}) as the operator-valued $\hbar$-power series, the zero order terms in $\hbar$-expansions provide the classical parameters $x_{AB}$. In particular, following 
the construction of the so-called "phenomenological Lagrangians" \cite{8Coleman,9}, one can distinguish two separate cases:
\begin{itemize}
\item[i)] if $x_{\mu\nu}=0$ and $x_{4\mu }=Mx_\mu\neq 0$ one obtains the Snyder models with preserved linear Lorentz covariance.
\item[ii)] in general case one can assume that $x_\mu\neq 0$ as well as $x_{\mu\nu}\neq 0$, what may lead to the spontaneous symmetry breaking of any part of the $D=4$ dS algebra $o(4,1)$\footnote{If we put $G=o(4,1)$, the coset decomposition $G=H\otimes \frac{G}{H}$ provides the construction of $H$-covariant phenomenological Lagrangians. If all symmetries of $G$ are spontaneously broken, one should put $H=$\textbf{1} and introduce in formula (\ref{Mcl-q}) 10 non-vanishing parameters $x_{AB}$ (see \ref{Mcl}).}.
\end{itemize}
If we introduce the vacuum state $|0\rangle$ with the lowest value of energy one can describe equivalently the relations (\ref{Mcl-q})-(\ref{Mcl}) as follows:
\begin{equation}
{M}_{AB}^{cl}|0\rangle=x_{AB}|0\rangle,\qquad \hat{M}_{AB}^{q}|0\rangle=0
\end{equation}
Because the parameters $x_{AB}$ are $\mathbb{C}$-numbers, one obtains that 
\begin{equation}
x_{AB}=\langle 0|\hat{M}_{AB}|0\rangle= (x_{\mu\nu}=-x_{\nu\mu},\, x_{4\mu}=-x_{\mu 4})
\end{equation}
and $x_{4\mu}=Mx_\mu$.

In the presence of SSB the vacuum (lowest energy state of a system) is not unique and we obtain the set of degenerate vacua, which may depend on the parameters $x_{AB}$. In such a case part of the symmetries providing the covariance of the Snyder model do not leave all vacua states invariant, what leads to the excitation of massless Goldstone bosons \cite{7Goldstone}. It should be added that in order to specify the set of SSB vacua one should employ the notion of a Hamiltonian, which can be calculated e.g. when Snyder algebra is derived from the Lagrangian formulation of the Snyder particle model (see e.g. \cite{33c,33d}).

\section{$\hbar$-perturbative solutions of spontaneously broken $D=4$ Snyder models}
Our aim in this paper is to employ the perturbative $\hbar$-expansions in order to consider the solutions of Snyder model with spontaneously broken $D=4$ dS symmetries. We expand the $o(4,1)$ generators $\hat{M}_{AB}$ in the $\hbar$-power series
\begin{equation}\label{Mab_exp}
\hat{M}_{AB}=M_{AB}^{(0)}+\hbar \hat{M}_{AB}^{(1)}+\hbar^2\hat{M}_{AB}^{(2)}+\ldots
\end{equation}
where $M_{AB}^{(0)}=M_{AB}^{cl}=x_{AB}$ describe the non-vanishing classical $\hbar$-independent part of the quantum generators $\hat{M}_{AB}$, what indicates the presence of SSB. The variables $x_{AB}$ (see (16)) can be treated as the Nambu-Goldstone (NG) coordinates which can be extended by adding dual NG momenta $p_{AB}=-p_{BA}$ and subsequently form the quantum phase space $(x_{AB},p_{AB})$ satisfying the tensorial canonical Heisenberg algebra\footnote{See also \cite{Nambu2004}, where the NG momenta for Lie algebras were considered.} 
\begin{equation}
\lbrack x_{AB},x_{CD}]=[p_{AB},p_{CD}]=0,\qquad \lbrack
x_{AB},p_{CD}]=i\hbar (\eta _{AC}\eta _{BD}-\eta _{AD}\eta _{BC}).
\label{tHalg}
\end{equation}
In order to calculate the $\hbar$-perturbative solutions of $D=4$ Snyder model, the presence of non-vanishing tensorial momenta is needed for the iterative explicit calculations of consecutive terms $\hat{M}_{AB}^{(n)}$ $(n=1,2,\ldots)$ in $\hbar$-power series (\ref{Mab_exp}). Such tensorial Heisenberg algebra (\ref{tHalg}) has been already postulated in earlier papers (see e.g. \cite{30,31}) but without exposing the relation to SSB and the perturbative solutions as $\hbar$-power series. It follows that $x_{AB}$ describe the Abelian order parameters which parametrize the set of possible spontaneously broken degenerate vacuum states $|0;x_{AB}\rangle$ if they satisfy the relations $\hat{M}_{AB}|0;x_{AB}\rangle=x_{AB}|0;x_{AB}\rangle$ and have the same minimal energy (Hamiltonian) eigenvalue.\\

\textit{$\alpha )$ perturbative $\hbar $-expansion: first order in $\hbar $}%
\newline

From relation \eqref{dSMM} one gets:
\begin{equation}  \label{xAB_MCD}
\left[ x_{AB}{},\hat{M}_{CD}{}^{\left( 1\right) }\right] +\left[ \hat{M}%
_{AB}{}^{\left( 1\right) },x_{CD}{}\right] =i(\eta _{AC}x_{BD}-\eta
_{AD}x_{BC}-\eta _{BC}x_{AD}+\eta _{BD}x_{AC})
\end{equation}
where $\eta_{AB}=diag(-1,1,1,1,1)$ and relations (\ref{Mx},\ref{snyder_m})
lead to
\begin{equation}  \label{xmn_Mrs}
\left[ x_{\mu\nu}{},\hat{M}_{\rho\sigma}{}^{\left( 1\right) }\right] -\left[ x_{\rho\sigma}{},\hat{M}
_{\mu\nu}{}^{\left( 1\right) }\right] =i(\eta _{\mu\rho}x_{\nu\sigma}-\eta
_{\mu\sigma}x_{\nu\rho}-\eta _{\nu\rho}x_{\mu\sigma}+\eta _{\nu\sigma}x_{\mu\rho}),
\end{equation}
\begin{equation}  \label{S_firstxM}
\left[ x_{\mu }{},\hat{M}_{\rho \sigma}{}^{\left( 1\right) }\right] -\left[
x_{\rho \sigma }{},\hat{x}_{\mu }{}^{\left( 1\right) }\right] =i(\eta _{\mu
\sigma }x_{\rho }-\eta _{\mu \rho }x_{\sigma }),\quad
\end{equation}
\begin{equation}  \label{S_firstxx}
\left[ x_{\mu }{},\hat{x}_{\nu }{}^{\left( 1\right) }\right] -\left[ x_{\nu
},\hat{x}_{\mu }{}^{\left( 1\right) }{}\right] =\frac{i}{M^{2}}x_{\mu \nu
}.\quad
\end{equation}
In order to solve the relations (\ref{xAB_MCD}-\ref{S_firstxx}) we employ the generalized momenta $p_{AB}=\left(p_{\mu \nu },p_{\mu }\right) $
(see (\ref{tHalg})). From (\ref{xmn_Mrs}) and (\ref{S_firstxM}) one
can obtain a  particular solution, given by\footnote{Subindex $S$ denotes
the Snyder case. In (\ref{solM1},\ref{solx1}) the factor $\hbar$ on the left hand side reflects the proportionality of quantum-mechanical momenta to $\hbar$.}
\begin{equation}\label{solM1}
\hbar\hat{M}_{\mu \nu ;S}^{\left( 1\right) }
=\frac{1}{2}\left(x_{\mu }^{~\ \rho }p_{\nu\rho }-x_{\nu }^{~\ \rho }p_{\mu\rho}\right)+
x_{\mu }p_{\nu }-x_\nu p_\mu
\end{equation}
and in consistency with (\ref{S_firstxx})
\begin{equation}\label{solx1}
\hbar\hat{x}_{\mu ;S}^{\left( 1\right) }=-\frac{1}{2M^{2}}x_{\mu \rho }p^{\rho }.
\end{equation}
The general first order solution depends on one free parameter and can be obtained by a suitable choice of 
 similarity transformations of the particular solutions (\ref{solM1},\ref{solx1}).

\bigskip \textit{$\beta )$ perturbative $\hbar $-expansion: second order in $%
\hbar $}
\newline

The second order counterpart of relation (\ref{xAB_MCD}) looks as follows:
\begin{equation}
\left[ x_{AB},\hat{M}_{CD}^{\left( 2\right) }\right] -\left[ x_{CD},\hat{M}%
_{AB}^{\left( 2\right) }\right] +\left[ \hat{M}_{AB}^{\left( 1\right) },\hat{%
M}_{CD}^{\left( 1\right) }\right] =i(\eta _{AC}\hat{M}_{BD}^{\left( 1\right)
}+\eta _{BD}\hat{M}_{AC}^{\left( 1\right) }-\eta _{BC}\hat{M}_{AD}^{\left(
1\right) }-\eta _{AD}\hat{M}_{BC}^{\left( 1\right) })
\end{equation}
which leads to:
\begin{equation}\label{sec_or1}
\left[ x_{\mu \nu },\hat{M}_{\rho \sigma }^{\left( 2\right) }\right] -\left[
x_{\rho \sigma },\hat{M}_{\mu \nu }^{\left( 2\right) }\right] =i(\eta _{\mu
\rho }\hat{M}_{\nu \sigma }^{\left( 1\right) }+\eta _{\nu \sigma }\hat{M}_{\mu
\rho }^{\left( 1\right) }-\eta _{\nu \rho }\hat{M}_{\mu \sigma }^{\left(
1\right) }-\eta _{\mu \sigma }\hat{M}_{\nu \rho }^{\left( 1\right) })-\left[
\hat{M}_{\mu \nu }^{\left( 1\right) },\hat{M}_{\rho \sigma }^{\left( 1\right) }
\right],
\end{equation}
\begin{equation}\label{sec_or2}
\left[ x_{\mu },\hat{M}_{\rho \sigma }^{\left( 2\right) }\right] -\left[
x_{\rho \sigma },\hat{x}_{\mu }^{\left( 2\right) }\right] =i(\eta _{\mu \sigma }\hat{x}_\rho^{(1)}-\eta _{\mu \rho}\hat{x}_\sigma^{(1)})-\left[ \hat{x}_{\mu }^{\left( 1\right)
},\hat{M}_{\rho \sigma }^{\left( 1\right) }\right],
\end{equation}
\begin{equation}\label{sec_or3}
\left[ x_{\mu },\hat{x}_{\sigma }^{\left( 2\right) }\right] -\left[ x_{\sigma },
\hat{x}_{\mu }^{\left( 2\right) }\right] =\frac{i}{M^{2}}\hat{M}_{\mu \sigma
}^{\left( 1\right) }-\left[ \hat{x}_{\mu }^{\left( 1\right) },\hat{x}_{\sigma
}^{\left( 1\right) }\right].
\end{equation}
Substituting in (\ref{sec_or1}-\ref{sec_or3}) the solutions (\ref{solM1},\ref{solx1}) one gets, in the second order of $\hbar$, the following particular solution:
\begin{equation}
\hbar^2 \hat{M}_{\mu \nu ;S}^{\left( 2\right) }=-\frac{1}{12}\left(x_{\mu \rho }
p^{\rho \sigma }p_{\nu \sigma }-x_{\nu \rho }p^{\rho \sigma }p_{\mu \sigma }
-2x^{\rho \sigma }p_{\mu \rho }p_{\nu \sigma }\right),
\label{solM2}
\end{equation}%
\begin{equation}
\hbar^2 \hat{x}_{\mu ;S}^{\left( 2\right) }=\frac{1}{M^2}\left(x_\rho p^\rho p_\mu +\frac{1}{4}\left( x_{\mu \rho }p_{\rho \sigma }
p_\sigma +x^{\rho \sigma }p_\rho p_{\mu \sigma }\right)\right).
\label{solx2}
\end{equation}%
General parameter-dependent solutions in the second $\hbar$-order can be obtained from the formulae (\ref{solM2},\ref{solx2}) by performing suitable similarity transformations.
One can also show that, in the perturbative $n$-th order of $\hbar$, the solutions $(\hat{x}_{\mu ;S}^{(n)}, \hat{M}_{\mu\nu ;S}^{(n)})$ are $n$-linear in momenta $p_{AB}=(p_{\mu\nu}, p_\mu)$.

\section{$\hbar$-perturbative solutions of spontaneously broken $D=4$ Yang models}

$D=4$ Yang model (see e.g. \cite{17}, \cite{Guo}-\cite{Manolakos}) is described by fifteen generators of $D=5$ dS algebra ${o}(5,1)$ ($K,L=0,1,2,3,4,5$)
\begin{equation}
\hat{M}_{KL}=\left( \hat{M}_{\mu \nu },\hat{M}_{4\mu }=M\hat{x}_{\mu },\hat{M%
}_{5\mu }=R\hat{q}_{\mu },\hat{M}_{45}=MR\hat{r}\right)   \label{MKL}
\end{equation}
which provides $D=4$ Lorentz-covariant quantum-deformed relativistic Heisenberg algebra with two deformation
parameters $\left( M,R\right) $ (of length dimensions $\left[ M\right]
=L^{-1},\left[ R\right] =L$) and dimensionless scalar Abelian generator $\hat{r}$, describing internal ${o}(2)$ symmetries.
 In general case, in Yang model one can introduce 
the following fifteen Abelian NG modes, which may break spontaneously all ${o}%
\left( 5,1\right) $ symmetries $\left( x_{KL}=-x_{LK}\right) $
\begin{equation}
x_{KL}=\left( x_{\mu \nu }, Mx_{\mu }, Rq_{\mu }, MRr\right) .  \label{var_x}
\end{equation}%
To solve the Yang model by using perturbative $\hbar $-expansion
one should also introduce fifteen canonically conjugated NG momenta
\begin{equation}
p_{KL}=\left( p_{\mu \nu },p_{\mu },k_{\mu },s\right)   \label{var_p}
\end{equation}
satisfying the canonical commutation relations \eqref{tHalg}, with the
additional relations
\begin{equation}
\left[ q_{\mu },k_{\nu }\right] =i\hbar \eta _{\mu \nu },\quad \quad \left[
r,s\right] =i\hbar .  \label{qk}
\end{equation}
Using the variables \eqref{var_x}, \eqref{var_p} one can obtain the $\hbar$-perturbative solutions of Yang model, which has been proposed in \cite{17} as the Lie-algebraic extension of the Snyder model \cite{16}. 
By
adding to ${o}\left( 4,1\right) $ {a fifth} space coordinate, one gets
the set of Lorentz-covariant formulae \eqref{MKL} where the generators $\left( \hat{x}_{\mu },\hat{q}_{\mu },\hat{r}\right) $ describe the quantum-deformed $D=4$
Heisenberg algebra, with one additional Abelian internal symmetry generator.

In Yang model one extends the Snyder relations \eqref{MM}, \eqref{Mx}, \eqref{snyder_m} by
the following set of algebraic equations
\begin{eqnarray}  \label{qq}
\left[ \hat{q}_{\mu },\hat{q}_{\nu }{}\right] &=&i\frac{\hbar }{R^{2}}\hat{M}%
_{\mu \nu }, \\
\left[ \hat{x}_{\mu }{},\hat{q}_{\nu }\right] &=&i\frac{\hbar }{MR}\eta
_{\mu \nu }\hat{r} \\
\left[ \hat{M}_{\mu \nu },\hat{q}_{\rho }{}\right] &=&i\hbar \left( \eta
_{\nu \rho }\hat{q}_{\mu }-\eta _{\mu \rho }\hat{q}_{\nu }\right) , \\
\left[ \hat{r},\hat{x}_{\mu }{}\right] &=& \frac{i\hbar}{M^2}\hat{q}_{\mu },
\\
\left[ \hat{r},\hat{q}_{\mu }{}\right] &=&-\frac{i\hbar}{R^2}\hat{x}_{\mu }
\label{rqhat}
\end{eqnarray}
{where $\hat{q}_{\mu }=q_{\mu }^{(cl)}+\hat{q}_{\mu }^{(q)},\quad \hat{r}=r^{(cl)}+\hat{r}^{(q)}$.}
The first order $\hbar $-approximation of the algebraic solutions of Yang model is obtained when in the relations \eqref{MM}, \eqref{Mx}, \eqref{snyder_m} and (\ref{qq}-\ref{rqhat}) we consider only the linear $\hbar $-terms. Besides (\ref{xmn_Mrs}-\ref{S_firstxx}) we get $\left( r\equiv
\hat{r}^{\left( 0\right) } =r^{(cl)}\right)$
\begin{eqnarray}  \label{qq0}
\left[ q_{\mu }{},\hat{q}_{\nu }{}^{\left( 1\right) }\right] -\left[ q_{\nu
}{},\hat{q}_{\mu }{}^{\left( 1\right) }\right] &=&\frac{i}{R^{2}}x_{\mu \nu
}, \\
\left[ x_{\mu }{},\hat{q}_{\nu }{}^{\left( 1\right) }\right] -\left[ q_{\nu
}{},\hat{x}_{\mu }{}^{\left( 1\right) }\right] &=&ir\eta_{\mu\nu},  \label{xq} \\
\left[ x_{\mu \nu }{},\hat{q}_{\rho }{}^{\left( 1\right) }\right] +\left[
\hat{M}_{\mu \nu }{}^{\left( 1\right) },q_{\rho }{}\right] &=&i\left( \eta
_{\nu \rho }q_{\mu }-\eta _{\mu \rho }q_{\nu }\right) ,  \label{xrhoM} \\
\left[ r{},\hat{x}_{\mu }{}^{\left( 1\right) }\right] +\left[ \hat{r}%
^{\left( 1\right) },x_{\mu }\right] &=&\frac{i}{M^2}q_{\mu },  \label{rx} \\
\left[ r{},\hat{q}_{\mu }{}^{\left( 1\right) }\right] +\left[ \hat{r}%
^{\left( 1\right) },q_{\mu }\right] &=&-\frac{i}{R^2}x_{\mu }.  \label{rq}
\end{eqnarray}
For the extended Snyder model, in the first order, we obtained the formulas \eqref{solM1}, \eqref{solx1}. In Yang model, due to the presence of additional
coordinates $(q_\mu,r)$ one should add their dual momenta $(k_\mu,s)$) (see \eqref{var_x},\eqref{var_p}) and extend the formulae \eqref{solM1}, \eqref{solx1} by terms which are linear in momenta $(k_\mu,s)$ (see \eqref{var_p}) as follows:
\begin{equation}
\hbar\hat{M}_{\mu \nu ;Y}^{\left( 1\right) }=\frac{1}{2}\left(x_{\mu }^{~\ \rho }p_{\nu\rho }-x_{\nu }^{~\ \rho }p_{\mu\rho}\right)+
x_{\mu }p_{\nu }-x_\nu p_\mu {-} q_{\mu }k_{\nu } {+} q_{\nu }k_{\mu } ,
\end{equation}
\begin{equation}\label{sol_x1Y}
\hbar\hat{x}_{\mu ;Y}^{\left( 1\right) }=-\frac{1}{2M^2}x_{\mu\rho}p^\rho +ax_{\mu\rho}k^{\rho}+brk_\mu+cq_\mu s,
\end{equation}
where $a,b,c$ are numerical constants. We add further the formulae:
\begin{equation}\label{sol_q1Y}
\hbar\hat{q}_{\mu }{}^{\left( 1\right) }=-\frac{1}{2R^{2}}x_{\mu \rho }{k}^{\rho }+\tilde{a}x_{\mu \rho }p^{\rho }+\tilde{b}rp_{\mu }+\tilde{c}x_\mu s,
\end{equation}
\begin{equation}\label{sol_r}
\hbar\hat{r}{}^{\left( 1\right) }=dq^\rho p_{\rho }+fx^\rho k_\rho,
\end{equation}
introducing additional five numerical constants $\tilde{a},\tilde{b},\tilde{c},d$ and $f$. The equations \eqref{qq0}-\eqref{rq} impose the following constraints on eight parameters in \eqref{sol_x1Y}-\eqref{sol_r}:
\begin{equation}
a+\tilde{a}=0,\qquad \tilde{b}=b+1, \qquad c-d=\frac{1}{M^2}, \qquad \tilde{c}-f=-\frac{1}{R^2}
\end{equation}
and imply the absence in formulae \eqref{sol_x1Y}-\eqref{sol_r} of the terms proportional to $p_{\mu\nu}$. We see therefore that the solutions of equations \eqref{qq0}-\eqref{rq} linear in $\hbar$ contain four unconstrained numerical parameters $a,b,c,f$.

The above calculation can be extended to higher orders in $\hbar$. In particular, in the second order of $\hbar$ we get the following set of algebraic equations (besides \eqref{sec_or1},\eqref{sec_or2},\eqref{sec_or3}):
\begin{eqnarray}\label{sec_or1Y}
\left[ q_{\nu }{},\hat{q}_{\sigma }{}^{\left( 2\right) }\right] -\left[
q_{\sigma }{},\hat{q}_{\nu }{}^{\left( 2\right) }\right] &=&\frac{i}{R^{2}}%
\hat{M}_{\nu \sigma }^{\left( 1\right) }-\left[ \hat{q}_{\nu }{}^{\left(
1\right) },\hat{q}_{\sigma }{}^{\left( 1\right) }\right], \\\label{sec_or2Y}
\left[ x_{\nu }{},\hat{q}_{\sigma }{}^{\left( 2\right) }\right] -%
\left[ q_{\sigma }{},\hat{x}_{\nu }{}^{\left( 2\right) }\right] &=&i\eta _{\nu
\sigma }\hat{r}^{\left( 1\right) }-\left[ \hat{x}_{\nu }{}^{\left( 1\right)
},\hat{q}_{\sigma }{}^{\left( 1\right) }\right], \\\label{sec_or3Y}
\left[ x_{\mu \nu }{},\hat{q}_{\sigma }{}^{\left( 2\right) }\right]
+\left[ \hat{M}_{\mu \nu }{}^{\left( 2\right) },q_{\sigma }{}\right] &=&i(\eta
_{\mu \sigma }\hat{q}_{\nu }^{\left( 1\right) }-\eta _{\nu \sigma }\hat{q}%
_{\mu }^{\left( 1\right) })-\left[ \hat{M}_{\mu \nu }{}^{\left( 1\right) },%
\hat{q}_{\sigma }{}^{\left( 1\right) }\right], \\
\label{sec_or4Y}
\left[ r{},\hat{x}_{\sigma }{}^{\left( 2\right) }\right] +\left[ 
\hat{r}{}^{\left( 2\right) },x_{\sigma }{}\right] &=&\frac{i}{M^{2}}\hat{q}%
_{\sigma }^{\left( 1\right) }-\left[ \hat{r}{}^{\left( 1\right) },\hat{x}%
_{\sigma }{}^{\left( 1\right) }\right], \\
\label{sec_or5Y}
\left[ r{},\hat{q}_{\sigma }{}^{\left( 2\right) }\right] +\left[ 
\hat{r}{}^{\left( 2\right) },q_{\sigma }{}\right] &=&-\frac{i}{R^{2}}\hat{x}%
_{\sigma }^{\left( 1\right) }-\left[ \hat{r}{}^{\left( 1\right) },\hat{q}%
_{\sigma }{}^{\left( 1\right) }\right].
\end{eqnarray}
where the second order terms $\hat{x}_\mu^{(2)}, \hat{q}_\mu^{(2)}, \hat{M}_{\mu\nu}^{(2)}, \hat{r}^{(2)}$ extending relations (29,30) are bilinear in momenta variables $p_{KL}$ (see (33)).

\section{Concluding remarks}
The main idea in the present paper is to show that in the algebraic models describing NC quantum space-times and  quantum phase spaces, the effects of SSB can also be present.

The non-commutative quantum space-times and quantum-deformed Heisenberg algebra representing NC phase spaces, both considered as the tools for the description of quantum gravity, are naturally expressed in the form of NC algebras, with the use of the framework of NC geometry.
We considered here simple and quite popular Snyder and Yang models, algebraically described by $o(4,1)$ and $o(5,1)$ Lie algebras, 
which have been often used in the current quantum gravity research, in particular exploiting NC quantum geometry (see \cite{15,19,47}). To be more specific, $o(4,1)$ describes an extended Snyder model, in which Snyder quantum space-time is a subspace of a larger non-commutative algebra, which includes Lorentz symmetry generators.
The algebraic basis of the extended Snyder
algebra is spanned by the generators defining Snyder quantum space-time coordinates $\hat{x}_\mu=\frac{1}{M}\hat{x}_{4\mu}$ as well as quantum tensorial coordinates $\hat{x}_{\mu\nu}=\hat{M}_{\mu\nu}$ describing the Lorentz algebra. Such models have been investigated mostly from the mathematical point of view 
by several authors (see e.g. \cite{25,27,29,30}), however the interpretation of the new Abelian tensorial coordinates (16) remained unclear, until the recent proposal \cite{33} of a link between  the Abelian tensorial coordinates and the modes providing SSB of Lorentz and de Sitter symmetries. 

In order to distinguish, in considered quantum algebraic models, the terms which provide SSB  we looked for the perturbative solutions expressed as the $\hbar$-power series ($\hbar$ is the Planck constant) for singling out the zero order $\hbar$-independent terms, which describe the classical parts of quantum solutions necessary for the appearance of SSB effects.
We add that above construction can be applied in supersymmetrized quantum
Snyder and Yang models (see e.g. \cite{49,50}), with SSB introducing fermionic Grassmanian NG degrees of freedom.

Finally, one can mention that the role of 
 spontaneously broken symmetries, background fields and classical modes in current quantum gravity research and the description of quantum Universe remains, at present, still an open issue. 
 
\section*{Acknowledgements}

J.~Lukierski and A. Pacho\l ~were supported by Polish NCN grant 2022/45/B/ST2/01067. A. Pacho\l\ 
acknowledges the support of the European Cooperation in Science and Technology COST Action CA18108. 



\end{document}